\documentclass[12pt, centerh1]{article}
\usepackage{amssymb,amsmath,colonequals,color,multirow,tikz}
\usepackage{natbib}
\usepackage{setspace}

\newcommand{\del}{\mathbf\Delta}

\newcommand{\vecy}{\mathbf{y}}

\newcommand{\vecx}{\mathbf{x}}
\newcommand{\vecw}{\mathbf{w}}

\newcommand{\vecX}{\mathbf{X}}
\newcommand{\vecU}{\mathbf{U}}

\newcommand{\vecz}{\mathbf{z}}

\newcommand{\vecpi}{\mbox{\boldmath$\pi$}}
\newcommand{\vectheta}{\mbox{\boldmath$\theta$}}

\newcommand{\varthet}{\mbox{\boldmath$\vartheta$}}
\newcommand{\vecmu}{\mbox{\boldmath$\mu$}}
\newcommand{\vecbeta}{\mbox{\boldmath$\beta$}}
\newcommand{\vecalpha}{\mbox{\boldmath$\alpha$}}
\newcommand{\mSigma}{\mbox{\boldmath$\Sigma$}}
\newcommand{\matsig}{\mbox{\boldmath$\Sigma$}}
\newcommand{\mDelta}{\mbox{\boldmath$\Delta$}}

\date{}
\begin{document}
\title{Mixtures of Variance-Gamma Distributions}
\author{Sharon M.\ McNicholas, Paul D.\ McNicholas\thanks{Department of Mathematics \& Statistics, McMaster University, Hamilton, Ontario, L8S~4L8, Canada. E-mail: mcnicholas@math.mcmaster.ca.} \ and Ryan P.\ Browne}
\date{}

\maketitle

\begin{abstract} 
A mixture of variance-gamma distributions is introduced and developed for model-based clustering and classification. The latest in a growing line of non-Gaussian mixture approaches to clustering and classification, the proposed mixture of variance-gamma distributions is a special case of the recently developed mixture of generalized hyperbolic distributions, and a restriction is required to ensure identifiability. Our mixture of variance-gamma distributions is perhaps the most useful such special case and, we will contend, may be more useful than the mixture of generalized hyperbolic distributions in some cases. In addition to being an alternative to the mixture of generalized hyperbolic distributions, our mixture of variance-gamma distributions serves as an alternative to the ubiquitous mixture of Gaussian distributions, which is a special case, as well as several non-Gaussian approaches, some of which are special cases. The mathematical development of our mixture of variance-gamma distributions model relies on its relationship with the generalized inverse Gaussian distribution; accordingly, the latter is reviewed before our mixture of variance-gamma distributions is presented. Parameter estimation carried out within the expectation-maximization framework.
\end{abstract}

\section{Introduction}
Finite mixture models assume that a population is a convex combination of a finite number of probability densities. Therefore, they represent a very natural framework for classification and clustering applications. We say that a random vector $\mathbf{X}$ arises from a (parametric) finite mixture distribution if, for all $\vecx \subset \mathbf{X}$, its density can be written $$f(\vecx\mid\varthet)= \sum_{g=1}^G \pi_g f_g(\vecx\mid\vectheta_g),$$ where $\pi_g >0$ such that $\sum_{g=1}^G \pi_g = 1$ are the mixing proportions, $f_1(\vecx\mid\vectheta_g),\ldots,f_G(\vecx\mid\vectheta_g)$ are the component densities, and $\varthet=(\vecpi,\vectheta_1,\ldots,\vectheta_G)$ denotes the vector of parameters with $\vecpi=(\pi_1,\ldots,\pi_G)$. The component densities $f_1(\vecx\mid\vectheta_1),\ldots,f_G(\vecx\mid\vectheta_G)$ are typically taken to be of the same type, most commonly multivariate Gaussian. Multivariate Gaussian mixtures have become increasingly popular in clustering and classification since they were first used for this purpose 50 years ago \citep{wolfe63}. The Gaussian mixture density is 
\begin{equation}\begin{split}\label{eqn:gmm}
f(\vecx\mid\varthet)= \sum_{g=1}^G \pi_g\phi(\vecx\mid\vecmu_g,\matsig_g),
\end{split}\end{equation}
where $\phi(\vecx\mid\vecmu_g,\matsig_g)$ is the multivariate Gaussian density with mean~$\vecmu_g$ and covariance matrix $\matsig_g$. When mixture models are used for clustering, the idiom `model-based clustering' is usually used. The term `model-based classification' is used similarly \citep[e.g.,][]{dean06,mcnicholas10c} and is synonymous with `partial classification' \citep[cf.][Section~2.7]{mclachlan92} --- both terms refer to a semi-supervised version of model-based clustering --- while model-based discriminant analysis is completely supervised \citep[cf.][]{hastie96}.

Suppose we observe data $\vecx_1,\ldots,\vecx_n$ and no component memberships are known, i.e., in a clustering scenario. Denote their component memberships by $z_{ig}$, where $z_{ig}=1$ if observation~$i$ is in component~$g$ and $z_{ig}=0$ otherwise. 
The likelihood for the Gaussian mixture model in this case can be written $$\mathcal{L(\varthet\mid\vecx)}=\prod_{i=1}^n\sum_{g=1}^G
\pi_g\phi(\vecx_i\mid\vecmu_g,\matsig_g).$$

Using the same notation, suppose that we are operating within the model-based classification paradigm; we have $n$ observations, of which $k$ have known group memberships. Following \cite{mcnicholas10c}, we order these $n$  observations so that the first $k$ have known group memberships; we incur no loss of generality as a result. Then, the likelihood can be written
\begin{equation}\label{eqn:mbclass}
\mathcal{L(\varthet\mid\vecx,\vecz)}=\prod_{i=1}^k\prod_{g=1}^G\left[\pi_g\phi(\vecx_i\mid\vecmu_g,\matsig_g)\right]^{z_{ig}}\prod_{j=k+1}^n\sum_{h=1}^{H}\pi_h\phi(\vecx_j\mid\vecmu_h,\matsig_h),
\end{equation}
for $H\geq G$. Note that this classification approach allows us to search for a number of components $H$ greater than the number of classes already observed, $G$. This gives an added flexibility in some respects but is usually ignored by taking $H=G$. From Equation~\ref{eqn:mbclass}, model-based clustering can be viewed as a special case of model-based classification that arises upon setting $k=0$ and $H=G$ within the classification paradigm \citep{mcnicholas10c,mcnicholas11b}. 

Finally, consider model-based discriminant analysis, and, again, order the $n$ observations so that the first $k$ have known group memberships. Now, we only use these first $k$ observations to estimate the unknown component memberships (as opposed to using all $n$), i.e., we form the likelihood 
$$\mathcal{L(\varthet\mid\vecx,\vecz)}=\prod_{i=1}^k\prod_{g=1}^G\left[\pi_g\phi(\vecx_i\mid\vecmu_g,\matsig_g)\right]^{z_{ig}}$$
and use the associated maximum likelihood estimates to compute the expected values
\begin{equation*}\label{eqn:zhat}
\hat{z}_{jg}\colonequals\frac{\hat\pi_g\phi(\vecx_j\mid\hat\vecmu_g,\hat\matsig_g)}
{\sum_{h=1}^{G}\hat\pi_h\phi(\vecx_j\mid\hat\vecmu_h,\hat\matsig_h)},
\end{equation*}
for $j=k+1,\ldots,n$. These expected values make up a discriminant rule and predicted group memberships are arrived at accordingly. Specifically, we use the maximum \textit{a posteriori} classifications $\text{MAP}\{\hat{z}_{jg}\}$, where $\text{MAP}\{\hat{z}_{jg}\}=1$ if max$_g\{ \hat{z}_{jg} \}$ occurs at component $g$, and $\text{MAP}\{\hat{z}_{jg}\}=0$ otherwise, for $j=k+1,\ldots,n$.
Note that it might be preferable to ignore known labels in some cases and treat the problem as a clustering problem rather than using model-based classification or discriminant analysis; whether one should do this depends \textit{inter alia} on the value of $k$ \citep[cf.][]{vrbik13b}.

Over the past few years, there has been a marked increase in work on non-Gaussian mixtures for clustering and classification. Initially, this work focused on mixtures of multivariate $t$-distributions \citep{peel00,andrews11b,andrews11a,andrews11c,andrews12,steane12}, which represent the most trivial departure from normality. More recently, work on the skew-normal distribution \citep{lin09} and the skew-$t$ distribution \citep{lin10,lee13,vrbik12,vrbik14,murray14a,murray14b} has been predominant, as well as work on other approaches \citep[e.g,][]{karlis07,browne11}. \cite{browne13c} add to the richness of this pallet by introducing a mixture of generalized hyperbolic distributions; many other approaches that have been tried are special cases of this `superclass'.

In Section~\ref{sec:method}, our methodology is developed drawing on connections with the generalized inverse Gaussian distribution. Parameter estimation is described (Section~\ref{sec:para}) and 
the paper concludes with a summary and suggestions for future work (Section~\ref{sec:conc}).

\section{Methodology}\label{sec:method}
\subsection{Generalized Inverse Gaussian Distribution}
The generalized inverse Gaussian (GIG) distribution was introduced by \cite{good53} and further developed by \cite{barndorff77}, \cite{blaesild78}, \cite{halgreen79}, and \cite{jorgensen82}. Write $Y\backsim\text{GIG}(\psi,\chi,\lambda)$ to indicate that a random variable~$Y$ follows a generalized inverse Gaussian (GIG) distribution with parameters $(\psi,\chi,\lambda)$ and density
\begin{equation}\label{gig}
p(y\mid\psi,\chi,\lambda) = \frac{\left({\psi}/{\chi}\right)^{\lambda/2}y^{\lambda-1}}{2K_\lambda\left( \sqrt{\psi \chi}\right)}\exp\left\{-\frac{\psi y+\chi/y}{2}\right\},
\vspace{-0.08in}
\end{equation}
for $y>0$, where $\psi,\chi\in\mathbb{R}^+$, $\lambda\in\mathbb{R}$, and $K_{\lambda}$ is the modified Bessel function of the third kind with index~$\lambda$. 

\subsection{Generalized Hyperbolic Distribution}\label{sec:ghd}
\cite{mcneil2005} give the density of a random variable $\vecX$ following the generalized hyperbolic distribution,
\begin{equation}\begin{split} \label{dist ghy1}
f(\vecx\mid\varthet) &= 
\left[ \frac{ \chi + \delta\left(\vecx, \vecmu\mid\matsig\right) }{ \psi+ \vecalpha'\matsig^{-1}\vecalpha} \right]^{\frac{\lambda-{p}/{2}}{2}}\\
&\qquad\qquad\qquad\qquad\times\frac{[{\psi}/{\chi}]^{{\lambda}/{2}}K_{\lambda - {p}/{2}}\left(\sqrt{[\psi+ \vecalpha'\matsig^{-1}\vecalpha] [ \chi +\delta(\vecx, \vecmu\mid\matsig)]}\right)}{ \left(2\pi\right)^{{p}/{2}} \left|\matsig\right|^{{1}/{2}} K_{\lambda}\left( \sqrt{\chi\psi}\right)\exp\left\{ \left( \vecmu- \vecx\right)'\matsig^{-1}\vecalpha\right\}},
\end{split}\end{equation}
where $\delta\left(\vecx, \vecmu\mid\matsig\right) = \left(\vecx - \vecmu\right)'\matsig^{-1}\left(\vecx - \vecmu\right)$  is the squared Mahalanobis distance between $\vecx$ and $\vecmu$, and $\varthet=\left(\lambda, \chi, \psi, \vecmu, \mDelta, \vecalpha\right)$ is the vector of parameters. Herein, we use the notation $\vecX\backsim\mathcal{G}_p\left(\lambda, \chi, \psi, \vecmu, \matsig, \alpha\right)$ to indicate that a $p$-dimensional random variable $\vecX$ has the generalized hyperbolic density in Equation~\ref{dist ghy1}. Note that we use $\matsig$ to denote the covariance because, in this parameterization, we need to hold $|\matsig|=1$ to ensure idenitifiability. 

A generalized hyperbolic random variable $\vecX$ can be generated by combining a random variable $Y\backsim\text{GIG}(\psi,\chi,\lambda)$ and a latent multivariate Gaussian random variable $\vecU\backsim\mathcal{N}(\mathbf{0},\matsig)$ using the relationship
\begin{equation} \label{dist of X given w}
\vecX = \vecmu+ Y \vecalpha + \sqrt{Y} \vecU,
\end{equation} 
and it follows that $\vecX\mid\left(Y = y \right) \backsim \mathcal{N}(\vecmu+y\vecalpha,y\matsig)$.
From Bayes' theorem, we can obtain $Y\mid(\vecX=\vecx)\backsim\text{GIG}(\psi+\vecalpha'\matsig^{-1}\vecalpha, \chi+\delta\left(\vecx,\vecmu\mid\matsig\right), \lambda-{p}/{2})$, cf.\ \cite{browne13c}.


\subsection{Variance-Gamma Distribution}\label{sec:vgamma}
The variance-gamma distribution arises as a special, limiting case of the generalized hyperbolic distribution (Equation~\ref{dist ghy1}) by setting $\lambda >0$ and $\chi \rightarrow 0$. Note that the variance-gamma distribution is also known as the generalized or Bessel function distribution. To obtain this representation of the variance-gamma distribution, we need to note that for small, positive $b$ we have 
$$K_a(b)\approx\begin{cases}
-\log\left(\frac{b}{2}\right)-\varepsilon & \text{if } a=0,\\
\frac{\Gamma(a)}{2}\left(\frac{2}{b}\right)^a & \text{if } a>0, 
\end{cases}$$
where $K_a(b)$ is the modified Bessel function of the third kind with index~$a$ and $\varepsilon$ is the Euler-Mascheroni constant. Noting that $\lambda>0$, we have
\begin{equation}\label{eqn:critical}
\left(\frac{\psi}{\chi}\right)^{\lambda/2}\frac{1}{K_{\lambda}(\sqrt{\chi\psi})}\approx\frac{2^{1-\lambda}}{\Gamma(\lambda)}\psi^{\lambda}
\end{equation}
for small, positive $\chi$. Using the result in Equation~\ref{eqn:critical}, we obtain the following variance-gamma density from the special, limiting case of Equation~\ref{dist ghy1}: 
\begin{equation}\begin{split} \label{eqn:vg1}
v(\vecx\mid\lambda, \psi, \vecmu, \matsig, \vecalpha) &= 
\left[\frac{\delta\left(\vecx,\vecmu\mid\matsig\right)}{\psi+ \vecalpha'\matsig^{-1}\vecalpha} \right]^{\frac{\lambda-{p}/{2}}{2}}\frac{2^{1-\lambda}\psi^{\lambda}K_{\lambda - {p}/{2}}\left(\sqrt{(\psi+ \vecalpha'\matsig^{-1}\vecalpha)\delta(\vecx, \vecmu\mid\matsig)}\right)}{\Gamma(\lambda)\left(2\pi\right)^{{p}/{2}} \left|\matsig\right|^{{1}/{2}} \exp\left\{\left(\vecmu-\vecx\right)'\matsig^{-1}\vecalpha\right\}},
\end{split}\end{equation}
for $\lambda>0$ and with the same notation as before. 
We use the notation $\vecX\backsim\mathcal{V}_p\left(\lambda, \psi, \vecmu, \matsig, \vecalpha\right)$ to indicate that a $p$-dimensional random variable $\vecX$ has the variance-gamma density in Equation~\ref{eqn:vg1}. 

By analogy with the generalized hyperbolic case, a random variable $\vecX\backsim\mathcal{V}_p\left(\lambda, \psi, \vecmu, \matsig, \vecalpha\right)$ can be generated by combining a random variable $Y\backsim\text{gamma}(\lambda,\psi/2)$ and a latent multivariate Gaussian random variable $\vecU\backsim\mathcal{N}(\mathbf{0},\matsig)$ using the relationship
\begin{equation} \label{dist of X given w}
\vecX = \vecmu+ Y \vecalpha + \sqrt{Y} \vecU.
\end{equation}
Note that $Y\backsim\text{gamma}(\lambda,\psi/2)$ denotes a gamma random variable $Y$ with the shape-rate parameterization, i.e., with density
\begin{equation} \label{eqn:gamma}
g(y\mid\lambda,\psi/2)=\frac{(\psi/2)^{\lambda}}{\Gamma(\lambda)}y^{\lambda-1}\exp\left\{-({\psi}/{2})y\right\},
\end{equation}
for $y>0$, $\lambda>0$, and $\psi>0$.
From Equation~\ref{dist of X given w}, we have $\vecX\mid\left(Y = y \right) \backsim \mathcal{N}(\vecmu+y\vecalpha,y\del)$. Therefore, noting that
\begin{equation*}\begin{split} \label{eqn:trick}
\delta(\vecx,\vecmu+y\vecalpha\mid y\matsig) &= \vecalpha'\matsig^{-1}\vecalpha y -(\vecx-\vecmu)'\matsig^{-1}\vecalpha-\vecalpha'\matsig^{-1} (\vecx-\vecmu)+\delta(\vecx,\vecmu\mid\matsig)/y,
\end{split}\end{equation*}
Bayes' theorem gives
\begin{equation*}\begin{split}
&f(y\mid\vecx)=\frac{\phi(\vecx\mid y)g(y)}{v(\vecx)}\\
&\qquad=\left[\frac{\psi+\vecalpha'\matsig^{-1}\vecalpha}{\delta\left(\vecx,\vecmu\mid\matsig\right)}\right]^{\frac{\lambda - {p}/{2}}{2}}
\frac{y^{\lambda-{p}/{2}-1}\exp\left\{-\left[y\left( \psi+ \vecalpha'\matsig^{-1}\vecalpha\right) + 
y^{-1}\delta\left(\vecx,\vecmu\mid\matsig\right)\right]/2\right\}}{2K_{\lambda-{p}/{2}} \left(\sqrt{\left(\psi+ \vecalpha'\matsig^{-1}\vecalpha\right)\delta\left(\vecx,\vecmu\mid\matsig\right)}\right)},
\end{split}\end{equation*}
and so 
$Y\mid(\vecX=\vecx)\backsim\text{GIG}(\psi+\vecalpha'\matsig^{-1}\vecalpha, \delta\left(\vecx,\vecmu\mid\matsig\right), \lambda-{p}/{2})$.

There are several limiting and special cases of the variance-gamma distribution. Most relevant to clustering applications is the special case called the asymmetric Laplace distribution \citep[cf.][]{kotz2001}, which arises upon setting $\lambda =1$ and $\psi= 2$. With a minor modification, the asymmetric Laplace distribution was recently used for mixture model-based clustering and classification by \cite{franczak14}.

Unfortunately, an identifiability issue will arise if we proceed with Equation~\ref{eqn:vg1} as is. To see why this is so, consider that Equation~\ref{eqn:vg1} can be written
\begin{equation*}\begin{split} \label{eqn:vg}
v(\vecx\mid\lambda, \psi, &\vecmu, \matsig, \vecalpha)\\& = 
\left[\left(\frac{1}{\psi^2}\right)\frac{\psi\delta\left(\vecx,\vecmu\mid\matsig\right)}{1+ \frac{1}{\psi}\vecalpha'\matsig^{-1}\vecalpha} \right]^{\frac{\lambda-{p}/{2}}{2}}\frac{2^{1-\lambda}\psi^{\lambda}K_{\lambda - {p}/{2}}\left(\sqrt{(1+ \frac{1}{\psi}\vecalpha'\matsig^{-1}\vecalpha)\psi\delta(\vecx, \vecmu\mid\matsig)}\right)}{\Gamma(\lambda)\left(2\pi\right)^{{p}/{2}} \left|\matsig\right|^{{1}/{2}} \exp\left\{\left(\vecmu-\vecx\right)'\matsig^{-1}\vecalpha\right\}},\\
&=\psi^{-\lambda+\frac{p}{2}}\left[\frac{\psi\delta\left(\vecx,\vecmu\mid\matsig\right)}{1+ \frac{1}{\psi}\vecalpha'\matsig^{-1}\vecalpha} \right]^{\frac{\lambda-{p}/{2}}{2}}\frac{2^{1-\lambda}\psi^{\lambda-\frac{p}{2}}K_{\lambda - {p}/{2}}\left(\sqrt{(1+ \frac{1}{\psi}\vecalpha'\matsig^{-1}\vecalpha)\psi\delta(\vecx, \vecmu\mid\matsig)}\right)}{\Gamma(\lambda)\left(2\pi\right)^{{p}/{2}} \psi^{-\frac{p}{2}}\left|\matsig\right|^{{1}/{2}} \exp\left\{\left(\vecmu-\vecx\right)'\matsig^{-1}\vecalpha\right\}},\\
&=\left[\frac{\delta\left(\vecx,\vecmu\mid\matsig_*\right)}{1+ \vecalpha_*'\matsig_*^{-1}\vecalpha_*} \right]^{\frac{\lambda-{p}/{2}}{2}}\frac{2^{1-\lambda}K_{\lambda - {p}/{2}}\left(\sqrt{(1+ \vecalpha_*'\matsig_*^{-1}\vecalpha_*)\delta(\vecx, \vecmu\mid\matsig_*)}\right)}{\Gamma(\lambda)\left(2\pi\right)^{{p}/{2}}\left|\matsig_*\right|^{{1}/{2}} \exp\left\{\left(\vecmu-\vecx\right)'\matsig_*^{-1}\vecalpha_*\right\}},\\
\end{split}\end{equation*}
where $$\vecalpha_*=\frac{1}{\psi}\vecalpha \qquad \text{and} \qquad \matsig_*^{-1}=\psi\matsig^{-1}.$$
To overcome this problem, set $\mathbb{E}[Y]=1$ in Equation~\ref{dist of X given w}, i.e., impose the restriction $\lambda=\psi/2$. For notational clarity, we define $\gamma\colonequals\lambda=\psi/2$. Then, our variance-gamma density can be written
\begin{equation}\begin{split} \label{eqn:vg2}
v_*(\vecx\mid\gamma, \vecmu, \matsig, \vecalpha) &= 
\left[\frac{\delta\left(\vecx,\vecmu\mid\matsig\right)}{2\gamma+ \vecalpha'\matsig^{-1}\vecalpha} \right]^{\frac{\gamma-{p}/{2}}{2}}\frac{2\gamma^{\gamma}K_{\gamma - {p}/{2}}\left(\sqrt{(2\gamma+ \vecalpha'\matsig^{-1}\vecalpha)\delta(\vecx, \vecmu\mid\matsig)}\right)}{\Gamma(\gamma)\left(2\pi\right)^{{p}/{2}} \left|\matsig\right|^{{1}/{2}} \exp\left\{\left(\vecmu-\vecx\right)'\matsig^{-1}\vecalpha\right\}},
\end{split}\end{equation}
and $Y\mid(\vecX=\vecx)\backsim\text{GIG}(2\gamma+\vecalpha'\matsig^{-1}\vecalpha, \delta\left(\vecx,\vecmu\mid\matsig\right), \gamma-{p}/{2})$.
Our mixture of variance-gamma distributions has density
$$v_{\text{{\tiny mix}}}(\vecx\mid\varthet)= \sum_{g=1}^G \pi_g v_*(\vecx\mid\gamma_g, \vecmu_g, \matsig_g, \vecalpha_g),$$ with the same notation as before, with subscript $g$ indexing over components.

\section{Parameter Estimation}\label{sec:para}

An expectation-maximization (EM) algorithm \citep{dempster77} is used for parameter estimation. The EM algorithm iteratively facilitates maximum likelihood estimation when data are incomplete, i.e., when there are missing and/or latent data, or are treated as being incomplete. For our mixture of variance-gamma distributions, the missing data consist of the unknown component membership labels and the latent variables $Y_{ig}$. We denote group membership labels by $z_{ig}$, where $z_{ig}=1$ if observation $i$ is in component~$g$ and $z_{ig}=0$ otherwise. The latent variables $Y_{ig}$ are assumed to follow gamma distributions (Equation~\ref{eqn:gamma}). For illustration, we assume a clustering paradigm in what follows; accordingly, none of the component membership labels are known. The complete-data log-likelihood for our mixture of variance-gamma distributions is given by
\begin{equation*}\begin{split} \label{mixture likelihood}
&l_c(\varthet\mid \vecx, \vecy,\mathbf{z}) 
= \sum_{i=1}^n\sum_{g=1}^G z_{ig}\bigg[\log\pi_g + \sum_{j=1}^p \log \phi\left(\vecx_i\mid\vecmu_g + y_{ig}\vecalpha_g , y_{ig}\mSigma_g \right)  + \log g(y_{ig}\mid\gamma_g,\gamma_g)\bigg]\\ 
&= C -\frac{1}{2}\sum_{i=1}^n\sum_{g=1}^Gz_{ig}\log\left|\mSigma_g^{-1}\right|+\sum_{i=1}^n\sum_{g=1}^Gz_{ig}\log g(y_{ig}\mid\gamma_g,\gamma_g)\\ 
& \ -\frac{1}{2}\mbox{tr}\Bigg\{\sum_{g=1}^G \mSigma_g^{-1} \sum_{i=1}^n z_{ig}\bigg[ \frac{1}{y_i}(\vecx_i-\vecmu_g)(\vecx_i-\vecmu_g)'-(\vecx_i-\vecmu_g)\vecalpha_g'-\vecalpha_g(\vecx_i-\vecmu_g)' + y_i\vecalpha\vecalpha'\bigg]\Bigg\}.
\end{split}\end{equation*}
where $C$ does not depend on the model parameters $\varthet=\left\{\pi_g, \gamma_g, \chi_g, \vecmu_g, \matsig_g, \vecalpha_g : g=1,\ldots,G\right\}$. 

In the E-step, the expected value of the complete-data log-likelihood is computed. In our case, or any other case where the model is from the exponential family, this is equivalent to replacing the sufficient statistics of the missing data by their expected values in the complete-data log-likelihood $l_c(\varthet\mid \vecx, \vecw,\mathbf{z})$. In our case, the missing data are the latent variables $Y_{ig}$ and the group membership labels~$Z_{ig}$. These two sources of missing data are independent; therefore, we only need the marginal conditional distribution for the latent variable and group memberships given the observed data. We will need following expectations:
\begin{equation*}\begin{split}
&\mathbb{E}\left[ Z_{ig} \mid \mathbf{x}_i \right] =  \frac{ \pi_g f(\mathbf{x}_i \mid\boldsymbol{\theta}_g)}{\sum_{h=1}^G \pi_h f(\mathbf{x}_i\mid\boldsymbol{\theta}_h)} \equalscolon \hat{z}_{ig},\\
&\mathbb{E}\left[ Y_{ig} \mid \vecx_i, z_{ig}=1 \right] = 
\sqrt{\frac{\delta\left(\vecx,\vecmu\mid\matsig\right)}{2\gamma+\vecalpha'\matsig^{-1}\vecalpha}}\frac{K_{\gamma-p/2+1}\left(\sqrt{(2\gamma+\vecalpha'\matsig^{-1}\vecalpha)\delta\left(\vecx,\vecmu\mid\matsig\right)}\right)}{K_{\gamma-p/2} \left(\sqrt{(2\gamma+\vecalpha'\matsig^{-1}\vecalpha)\delta\left(\vecx,\vecmu\mid\matsig\right)}\right)} \equalscolon a_{ig},\\
&\mathbb{E}\left[{1}/{Y_{ig}} \mid \vecx_i, z_{ig}=1 \right] = \sqrt{\frac{2\gamma+\vecalpha'\matsig^{-1}\vecalpha}{\delta\left(\vecx,\vecmu\mid\matsig\right)}}\frac{K_{\gamma-p/2+1}\left(\sqrt{(2\gamma+\vecalpha'\matsig^{-1}\vecalpha)\delta\left(\vecx,\vecmu\mid\matsig\right)}\right)}{K_{\gamma-p/2}\left(\sqrt{(2\gamma+\vecalpha'\matsig^{-1}\vecalpha)\delta\left(\vecx,\vecmu\mid\matsig\right)}\right)}\\&\qquad\qquad\qquad\qquad\qquad\qquad\qquad\qquad\quad\qquad\qquad\qquad\quad\qquad\qquad - \frac{2(\gamma-p/2)}{\delta\left(\vecx,\vecmu\mid\matsig\right)}\equalscolon b_{ig},\\
&\mathbb{E}[\mbox{log}(Y_{ig})\mid\vecx_i, z_{ig}=1] =\log\sqrt{\frac{\delta\left(\vecx,\vecmu\mid\matsig\right)}{2\gamma+\vecalpha'\matsig^{-1}\vecalpha}}+\Bigg[\frac{1}{K_{\gamma-p/2} \left(\sqrt{(2\gamma+\vecalpha'\matsig^{-1}\vecalpha)\delta\left(\vecx,\vecmu\mid\matsig\right)}\right)}\\&\qquad\qquad\qquad\qquad\qquad\qquad\qquad\times\frac{\partial}{\partial\lambda}K_{\gamma-p/2}\left(\sqrt{(2\gamma+\vecalpha'\matsig^{-1}\vecalpha)\delta\left(\vecx,\vecmu\mid\matsig\right)}\right)\Bigg]
\equalscolon c_{ig}.
\end{split}\end{equation*}
Note that the relatively tractable forms of $\mathbb{E}\left[ Y_{ig} \mid \vecx_i, z_{ig}=1 \right]$, $\mathbb{E}\left[{1}/{Y_{ig}} \mid \vecx_i, z_{ig}=1 \right]$, and $\mathbb{E}[\mbox{log}(Y_{ig})\mid\vecx_i, z_{ig}=1]$ arise because $Y_{ig}~|~\vecx$ is GIG (cf.\ Section~\ref{sec:vgamma}).

In the M-step, we maximize the expected value of the complete-data log-likelihood to obtain updates for our parameters. The updates for $\pi_g$, $\vecmu_g$, and $\vecbeta_g$ are 
\begin{equation*}\begin{split}
\hat{\pi}_g=\frac{n_g}{n},\qquad\hat{\vecmu}_g = \frac{ \sum_{i=1}^n \vecx_i\hat{z}_{ig}( \overline{a}_g b_{ig}-1)}{\sum_{i=1}^n \hat{z}_{ig} (\overline{a}_gb_{ig}-1)}, \qquad\text{and}\qquad
\hat{\vecbeta}_g = \frac{ \sum_{i=1}^n\vecx_i \hat{z}_{ig}(b_{ig}-\overline{b}_{g})}{\sum_{i=1}^n \hat{z}_{ig} (\overline{a}_g b_{ig}-1)},
\end{split}\end{equation*}
respectively, where $n_g= \sum_{i=1}^n\hat{z}_{ig}$, $\overline{a}_g = ({1}/{n_g})\sum_{i=1}^n\hat{z}_{ig} a_i$, and $\overline{b}_g = ({1}/{n_g})\sum_{i=1}^n\hat{z}_{ig} b_i$. The update for $\matsig_g$ is given by
\begin{equation} \label{update sigma}
\hat{\matsig}_g =\frac{1}{n_g}\sum_{i=1}^n\hat{z}_{ig} b_{ig} (\vecx_i - \hat{\vecmu}_g)  (\vecx_i - \hat{\vecmu}_g)'-\hat{\vecalpha}_g\left( \overline{\vecx}_g- \hat{\vecmu}_g \right)'  - \left( \overline{\vecx}_g- \hat{\vecmu}_g \right) (\hat{\vecalpha}_g)' + \overline{a}_g \hat{\vecalpha}_g (\hat{\vecalpha}_g)',
\end{equation}
where $ \overline{\vecx}_g=({1}/{n_g})\sum_{i=1}^n\hat{z}_{ig}\vecx_i$. 
The update for $\gamma_g$ arises as the solution to the equation
\begin{equation*}
\sum^n_{i=1}\hat{z}_{ig}\varphi({\gamma}_g)-\sum^n_{i=1}\hat{z}_{ig}\log\gamma_{g} = \sum^n_{i=1}\hat{z}_{ig}c_{ig}-\sum^n_{i=1}\hat{z}_{ig} a_{ig}+n_g,
\end{equation*}
where $\varphi(\cdot)$ is the digamma function.

We determine convergence of our EM algorithms via the Aitken acceleration \citep{aitken26}. Specifically, the Aitken acceleration is used to estimate the asymptotic maximum of the log-likelihood at each iteration and we consider the algorithm converged if this estimate is sufficiently close to the current log-likelihood. The Aitken acceleration at iteration $k$ is 
$$a^{(k)}=\frac{l^{(k+1)}-l^{(k)}}{l^{(k)}-l^{(k-1)}},$$
where $l^{(k)}$ is the log-likelihood at iteration $k$. An asymptotic estimate of the log-likelihood at iteration $k+1$ is
$$l_{\infty}^{(k+1)}=l^{(k)}+\frac{1}{1-a^{(k)}}(l^{(k+1)}-l^{(k)}),$$
and the algorithm can be considered to have converged when
$l_{\infty}^{(k)}-l^{(k)}<\epsilon$ \citep{bohning94,lindsay95}.

We use the Bayesian information criterion \citep[BIC;][]{schwarz1978} to determine the number of components $G$ when unknown. The BIC can be written $$\text{BIC} = 2l(\mathbf{x}, \hat{\boldsymbol{\theta}}) - \rho \log n,$$ where $l(\mathbf{x}, \hat{\boldsymbol{\theta}})$ is the maximized log-likelihood, $\hat{\boldsymbol{\theta}}$ is the maximum likelihood estimate of ${\boldsymbol{\theta}}$, $\rho$ is the number of free parameters in the model, and $n$ is the number of observations. The BIC has long been used for mixture model selection and its use was motivated through Bayes factors \citep{kass95,kass95b,dasgupta98}. While many alternatives have been suggested \citep[e.g.,][]{biernacki00} none have proved superior.

\section{Discussion}\label{sec:conc}
We have introduced a mixture of variance-gamma distributions, where component density arises as a restricted ($\lambda=\gamma, \psi=2\gamma$), special, limiting case of the generalized hyperbolic distribution. Updates are derived for parameter estimation within the EM algorithm framework, which is made feasible by exploitation of the relationship with the GIG distribution. Figure~\ref{fig:flow} is an attempt to place our approach within the mosaic of recent work on mixture modelling approaches with component densities that follow the generalized hyperbolic distribution or a variant thereof \citep[e.g.,][]{browne13c,franczak14,murray14a}. Clearly, our variance-gamma mixtures, with components denoted ``Variance-Gamma$^*$" in Figure~\ref{fig:flow}, have a role to play and it will be interesting to investigate their performance in applications. Implementing an EM algorithm to facilitate such applications is a subject of ongoing work. These applications will include model-based classification and discriminant analysis in addition to model-based clustering. The fractionally-supervised paradigm \citep{vrbik13b} will also be investigated. Finally, alternatives to the EM for parameter estimation will be investigated, including a variational Bayes approach along the lines of work by \cite{subedi14}.
%
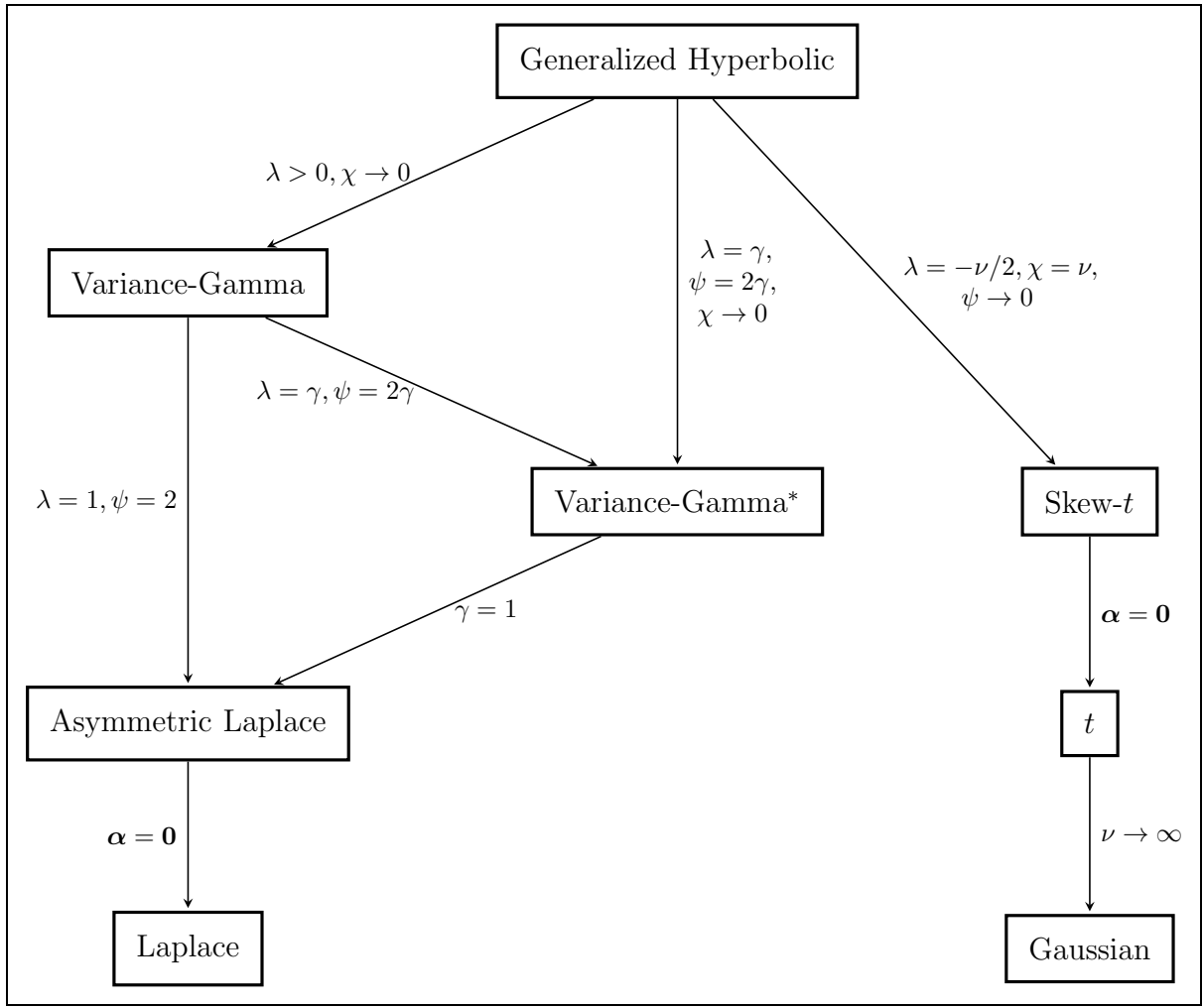
\begin{figure}[!h] 
\centering
\fbox{\centering\begin{tikzpicture}[
  every matrix/.style={ampersand replacement=\&,column sep=1.95cm,row sep=2cm},
  datastore/.style={draw,very thick,shape=rectangle,inner sep=.3cm},
  to/.style={->,>=stealth,shorten >=1pt,semithick,font=\footnotesize},
  every node/.style={align=center}]

\matrix{  
\&    \node[datastore] (gh) {Generalized Hyperbolic}; \& \\
\node[datastore] (vg) {Variance-Gamma}; \&   \&   \\
\& \node[datastore] (vgs) {Variance-Gamma$^*$};  \& \node[datastore] (st) {Skew-$t$};  \\
\node[datastore] (al) {Asymmetric Laplace}; \&  \& \node[datastore] (t) {$t$};  \\
\node[datastore] (l) {Laplace}; \&   \& \node[datastore] (g) {Gaussian};  \\
  };

  \draw[to] (gh) -- node[midway,left] {$\lambda>0, \chi\rightarrow{0}\;$} (vg);
  \draw[to] (gh) -- node[midway,right] {$\;\lambda=-\nu/2, \chi=\nu,$\\ $\psi\rightarrow{0}$} (st);
  \draw[to] (gh) -- node[midway,right] {$\;\lambda=\gamma,$\\ $\psi=2\gamma,$\\ $\chi\rightarrow{0}$} (vgs);
  \draw[to] (vg) -- node[midway,left] {$\lambda=\gamma,\psi=2\gamma\;$} (vgs);
  \draw[to] (vgs) -- node[midway,right] {$\;\gamma=1$} (al);
  \draw[to] (vg) -- node[midway,left] {$\lambda=1, \psi=2$} (al);
  \draw[to] (st) -- node[midway,right] {$\vecalpha=\mathbf{0}$} (t);
  \draw[to] (al) -- node[midway,left] {$\vecalpha=\mathbf{0}$} (l);
  \draw[to] (t) -- node[midway,right] {$\nu\rightarrow\infty$} (g);
 \end{tikzpicture}}
\caption{Some distributions available as special and limiting cases of the generalized hyperbolic distribution (Equation~\ref{dist ghy1}).}
\end{figure}\label{fig:flow}

\end{document}